\documentstyle[psfig]{mn}

\def\gs{\mathrel{\raise0.35ex\hbox{$\scriptstyle >$}\kern-0.6em
\lower0.40ex\hbox{{$\scriptstyle \sim$}}}}
\def\ls{\mathrel{\raise0.35ex\hbox{$\scriptstyle <$}\kern-0.6em
\lower0.40ex\hbox{{$\scriptstyle \sim$}}}}
\def\ls{\mathrel{\hbox{\rlap{\hbox{\lower4pt\hbox{$\sim$}}}\hbox{$<$}}}}
\def\gs{\mathrel{\hbox{\rlap{\hbox{\lower4pt\hbox{$\sim$}}}\hbox{$>$}}}}
\def\ergs {{\rm erg} \, {\rm s}^{-1}}
\def\mnras {{\sc MNRAS}}
\def\apj {ApJ}

\def\apjl {ApJL}
\def\aj {AJ}
\def\aap {A\&A}
\def\aaps {A\&AS}
\def\pasp {PASP}
\def\nat {Nature}

\title[Inter-cluster FOG Programme: Pilot Study Results]
      {Inter-cluster Filaments of Galaxies Programme: Pilot Study Survey 
and Results}
\author[K.\,A.\ Pimbblet and M.\ J.\ Drinkwater]
       {Kevin A.\ Pimbblet and Michael J.\ Drinkwater
        \vspace*{1mm}\\
        Department of Physics, University of Queensland, Brisbane,
        4072 Queensland, Australia\\
        pimbblet@physics.uq.edu.au}

\date{\fbox{\sc Draft: \today\ --- Do Not Distribute}}

\pagerange{000--000}

\begin{document}

\maketitle

\begin{abstract}
We present results from a pilot study of a new wide-field, multicolour (BVR)
CCD imaging project, designed to examine galaxy evolution along 
large scale filaments that connect clusters of galaxies at 
intermediate redshifts ($0.08 < z < 0.20$).
Our pilot dataset is based on 0.56 degree$^2$ of observations 
targeted on Abell~1079 and Abell~1084 using the Wide Field Imager 
on the Anglo-Australian Telescope.
We describe our data reduction pipeline and show that our 
photometric error is 0.04 mags.
By selecting galaxies that lie on the colour-magnitude relation
of the two clusters we verify the existence of a low density ($\sim 3$--$4$
Mpc$^{-2}$) filament population conjoining them at a distance of
$> 3 r_{Abell}$ from either cluster.
By applying a simple field correction, we characterize 
this filament population by examining their
colour distribution on a $(V-R)$--$(B-V)$ plane.
We confirm the galaxian filament detection at a $7.5 \sigma$ level 
using a cut at $M_V=-18$ 
and we discuss their broad properties.
\end{abstract}


\section{Introduction}

Large scale surveys of the Universe 
such as the 2dF Galaxy Redshift Survey (2dFGRS; e.g.\ Colless et al.\ 2001)
and the Sloan Digital Sky Survey (SDSS; e.g.\ Stoughton et al.\ 2002) 
clearly demonstrate the filamentary organization of galaxies.
These filaments of galaxies stretch between clusters
and superclusters of galaxies at all redshifts
(e.g.\ Kaldare et al.\ 2003; Einasto et al.\ 1997), 
forming a characteristic sponge-like structure through the 
Universe (Drinkwater 2000 and references therein).
Reproduction of such filaments is readily achieved in
modern hydrodynamical simulations of the Universe  
within the framework of cold dark matter (e.g.\ Jenkins et al.\ 1998).

Whilst much effort has been expended on analysing the nodes of these filaments
(i.e.\ clusters of galaxies; e.g.\ De Propris et al.\ 2002), 
only a little progress has been made into the largely over-looked large scale
structure beyond the core of these 
nodes (Pimbblet et al.\ 2002; Kodama et al.\ 2001 amongst others).

%
%
\begin{table*}
\begin{center}
\caption{Details of the global properties of the two 
clusters in the In-FOG-Pro pilot study.
For each cluster we give the coordinates of the cluster's X-ray
centre, the redshift ($z$) and its X-ray luminosity, in the
0.1--2.4\,keV passband (Pimbblet 2001; Ebeling et al.\ 1996; Edge, priv.\ comm.).
\hfil}
\begin{tabular}{lcccc}
\noalign{\medskip}
\hline
Cluster & R.A.\ & Dec.\ & $z$ & $L_X$ \\
         &  \multispan2{\hfil(J2000)\hfil} & & ($\times 10^{44} \ergs$) \\
\hline
Abell 1079 & 10 43 24.90 & $-$07 22 45 & 0.133 & $<$0.45 \\
Abell 1084 & 10 44 30.72 & $-$07 05 02 & 0.134 &  7.42 \\
\hline
\noalign{\smallskip}
\end{tabular}
  \label{tab:clusters}
\end{center}
\end{table*}

The Inter-cluster Filaments of Galaxies Programme (In-FOG-Pro)
is a long-term project to study a sample of large-scale structure
at intermediate redshifts ($0.07 \leq z \leq 0.20$) 
in the southern hemisphere.
Our broad goals are to understand evolution of filaments of galaxies and to
examine the role that filaments play in the
evolution of the (super-)clusters that they (intra-)connect.

For this, the pilot study, a potential filament was 
selected merely on the basis of observational availability 
during service time operations at the Anglo-Australian Telescope (AAT).
The filament selection technique for the larger 
programme will be presented in a future publication.
The simple premise is that the two target clusters of galaxies
observed, Abell~1079 and Abell~1084, 
are close to each other in redshift ($z\sim0.133$; Pimbblet
et al.\ 2002) and have a small spatial offset ($\sim 30$ arcmin), hence
they are expected to demonstrate a filamentary connection.
The global properties for the cluster targets in our pilot study 
are presented in Table~\ref{tab:clusters}.
To map their connecting filament, the Wide Field Imager (WFI) instrument 
is used as it provides a near-unrivalled field of view of 8 k pixels$^2$ 
which corresponds
to $\sim 6\ \rm{Mpc}^{2}$ at the redshift of our target 
filament\footnote{Throughout this
work, values of $H_0 = 50 $ km s$^{-1}$ Mpc$^{-1}$ and 
$q_o=0.5$ are adopted.}.

This article presents the details of the photometric reduction, 
calibration and analysis of the WFI imaging of the pilot study.
The plan of the paper is as follows: In \S2 we discuss the imaging
observations obtained for the pilot study, the data reduction methods,
catalogue construction and calibration.  
In \S3, after examining our internal consistency, we 
compare the galaxy catalogue derived from our WFI imaging 
with that from the Las Campanas/AAT
Rich Cluster Survery (LARCS; e.g.\ Pimbblet et al.\ 2001).
In \S4, we verify the existence of an inter-cluster filament
and by employing a simple field correction technique, we make 
a colour-colour analysis of these filament galaxies
and calculate the significane of the filament overdensity in \S5.
We summarize our main conclusions in \S6.

\section{Optical Imaging and Data Reduction}

The observations for our pilot study were
made and completed using WFI at the AAT in February 2003.
These data are presented in Table~\ref{tab:obs}.
Our prime observations are the ones taken on 01 February.  Since
they are taken in non-photometric conditions, 
however, necessary calibration observations
also need to be employed (02 February).

\begin{figure}
\centerline{\psfig{file=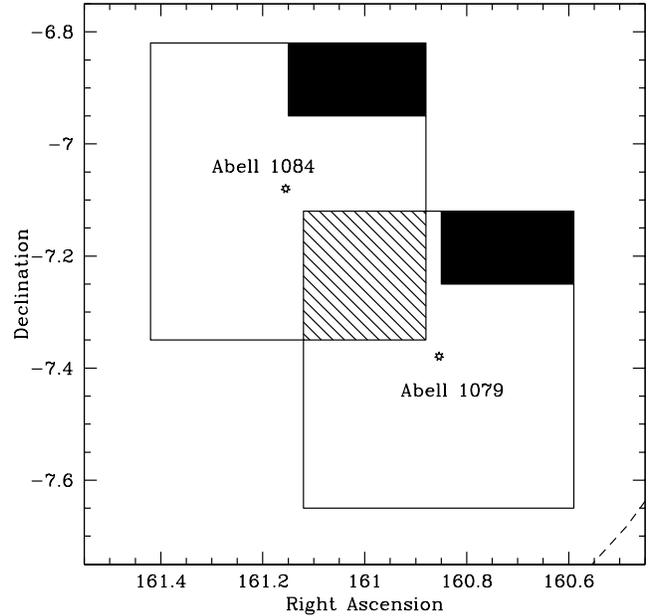,angle=0,width=3.5in}}
  \caption{\small{Schematic illustration of the observational layout used.
The stars mark the centres of the observations and approximate to
the cluster centres.  The solid regions correspond to CCD1 which 
is not used due to fringing.  The central hatched area is the
internal overlap whilst the dashed line (lower right corner) denotes the
approximate spatial extent of the LARCS catalogue.
The gaps between each CCD (not displayed) are $\sim 70$ pixels wide
and therefore any sources falling in these gaps are numerically
negligible compared to the total area surveyed.
}}
  \label{fig:layout}
\end{figure}

WFI is a mosaic imager that consists of 8 individual CCDs, each
2 k $\times$ 4 k pixels$^2$ in size.
This gives a pixel scale of $0.2295''$/pixel and a total mosaic field of
view for each exposure of $32.2' \times 31.7'$.
By targetting the observations on the cluster centres, the exposures
cover the filament and overlap by some 200 square 
arcmin (Figure~\ref{fig:layout}).

%
%
\begin{table}
\begin{center}
\caption{Log of AAT/WFI imaging observations made for the filament
of galaxies used in this work. The column headed `Passbands'
indicates the broad band filters used to observe each target cluster.
The $T_{Exp}$ column gives the exposure time whilst `Phot?' 
denotes whether the observation is made in photometric conditions
or not. \hfil}
\begin{tabular}{lccccc}
\noalign{\medskip}
\hline
Cluster & Date & Pass- & Seeing & $T_{Exp}$ & Phot? \\
     &    (2003) &  band    &   ($''$) & (sec)   & \\
\hline
Abell~1079 & Feb 01 & $B$ & 2.3--2.5 & 600 & no \\
           & Feb 01 & $V$ & 2.3--2.5 & 600 & no \\
           & Feb 01 & $R$ & 2.3--2.5 & 600 & no \\
           & Feb 02 & $B$ & 4.3--4.5 & 300 & yes \\
           & Feb 02 & $V$ & 4.3--4.5 & 300 & yes \\
           & Feb 02 & $R$ & 4.3--4.5 & 300 & yes \\
Abell~1084 & Feb 01 & $B$ & 2.3--2.5 & 600 & partial \\
           & Feb 01 & $V$ & 2.3--2.5 & 600 & no \\
           & Feb 01 & $R$ & 2.3--2.5 & 600 & no \\
           & Feb 02 & $B$ & 4.3--4.5 & 300 & yes \\
           & Feb 02 & $V$ & 4.3--4.5 & 300 & yes \\
           & Feb 02 & $R$ & 4.3--4.5 & 300 & yes \\
\noalign{\smallskip}  \hline
\end{tabular}
  \label{tab:obs}
\end{center}
\end{table}

%
%

These imaging data are reduced using standard packages within
{\sc iraf}, inparticular the {\sc mscred} package.
Steps undertaken initially include debiasing and 
preliminary flatfielding with twilight flats and creation
of super-flats from independent stacks of the science 
observations and the application of these to the
science pointings.

To test the precision of the flatfielding, the {\sc iraf} task
{\sc imstat} is used to obtain the mean background levels at
various locations on each mosaic tile.
The variation in the sky background is found to be $\leq 1$ percent
in all cases apart from CCD1.
CCD1 shows significant fringing patterns across most 
($\sim 70$ per cent) of the chip,
with the $R$-band imaging being the worst affected.  
Since the observations are only two times one-shot exposures 
per cluster (Table~\ref{tab:obs}), removal of the fringing
pattern proves to be an intractable problem. 
Unavoidably, therefore, our 
analysis herein discards CCD chip 1; see Figure~\ref{fig:layout}.

\subsection{Catalogue Construction and Calibration}

SExtractor (Bertin \& Arnouts 1996) is used to automatically
analyse the observations, detect sources and parametrize them.  
We catalogue
the $R$-band frames, detecting sources having more than 12 contiguous
pixels each $3\sigma$ of the sky above the background.
Since our $V$- and $B$-band observations are pointed in the
same direction (to better than 1 pixel accuracy)
and possess the same seeing quality (checked by measuring
the FWHM of stars in the field using SExtractor) 
as the $R$-band observation, we run SExtractor in
dual mode (same apertures) to derive colour information 
for these sources.

The photometric observations (02 February) are calibrated to 
observations of standard stars selected from 
Landolt (1992) scattered throughout the night at various airmasses.
Colour terms are fitted for by using standard star observations
taken at the same airmass.
Extinction is accounted for by fitting the variation in standard
star photometry in one night with their measured airmass.
In turn, these are used to calibrate the primary, partial- and non-photometric
science observations (01 February) by comparing magnitudes in
seeing-scaled apertures.
Our final zeropoint error is calculated to be $\sim 0.04$ mags in all
three passbands.

As all of our observations are one-shot images, cosmic ray rejection
is problematical.  We use the method of Rhoads (2000) which 
searches for features with significant power at spatial frequencies 
too high to be legitimate objects.  This eliminates all but the largest
of cosmic rays and satellite trails
(i.e.\ $>$ several 10s of contiguous pixels) which
are identified and removed by hand and/or SExtractor's {\sc flag} parameter.

Astrometry is performed individually on each CCD of the mosaic.
Approximately 50 $R<18.0$ stars per CCD are tied to the
positions from the APM (e.g.\ Maddox et al.\ 1990; 
http://www.ast.cam.ac.uk/apm/).
The positions of the unknown sources (both galaxies and stars) are
subsequently determined by use of the {\sc astrom} package
from {\sc starlink}.

\subsection{Star-Galaxy Separation}

%
%
\begin{figure}
\centerline{\psfig{file=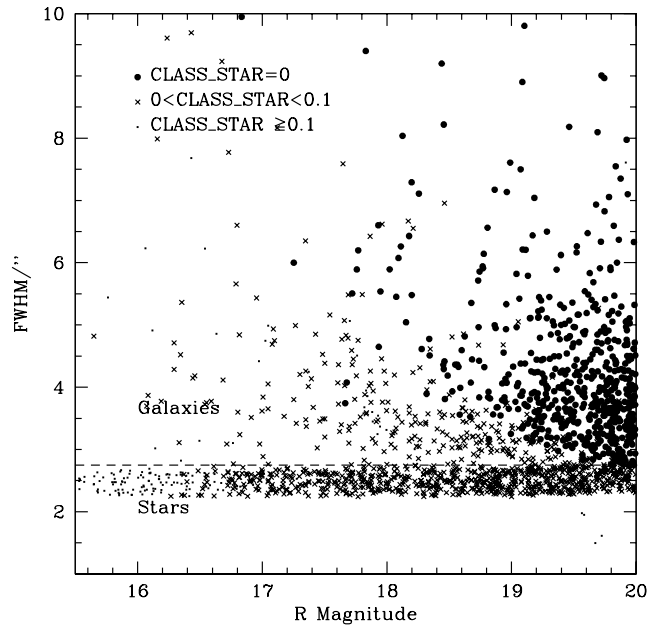,angle=0,width=3.5in}}
  \caption{\small{FWHM plotted as a function of $R$-band magnitude
for all objects in the catalogue.
Those objects that are {\it definitely} galaxies (i.e.\ {\sc class\_star}$=0$)
are emphasized as filled circles.  
Intermediate sources ($0 <  {\sc class\_star}
< 0.1$) are denoted by crosses.
The dashed horizontal line corresponds to {\sc fwhm}$=2.75''$ and effectively
differentiates stars (the clear locus of points below this line) 
from galaxies.
}}
  \label{fig:fwhm}
\end{figure}

To reliably differentiate between stars and galaxies in our
catalogue, we broadly follow the method of Pimbblet et al.\ (2001).
On a plane of magnitude versus {\sc fwhm}, stars should trace 
a locus of fixed {\sc fwhm} ($\sim$ seeing) at bright ($R<19.5$) 
magnitudes (Pimbblet et al.\ 2001).
Figure~\ref{fig:fwhm} displays this information for our catalogue.  
By employing a cut of {\sc fwhm}$>2.75''$, we are readily 
able to select a galaxy population from the parent catalogue.
When combined with a reasonable cut in SExtractor's {\sc class\_star} parameter
of $<0.1$ (see Figure~\ref{fig:fwhm}), 
we estimate that our ultimate stellar contamination 
rate is $<3$ per cent at bright magnitudes ($R<19.5$).
Although it is appropriate to consider this cut to be conservative, 
it still includes more compact 
galaxies (Pimbblet et al.\ 2001) but risks missing out
the newly discovered class of ultra-compact dwarf galaxies whose appearance 
is markedly stellar (Drinkwater et al.\ 2003).
To check that early-type galaxies have not been
classified as stars, we construct colour-magnitude diagrams
for those objects with {\sc class\_star}$>0.1$ and {\sc fwhm}$<2.75''$ 
and find that there is no hint of a colour-magnitude relation
(Visvanathan \& Sandage 1977).

\section{Photometric Accuracy of the Catalogue}

%
%
\begin{figure}
\centerline{\psfig{file=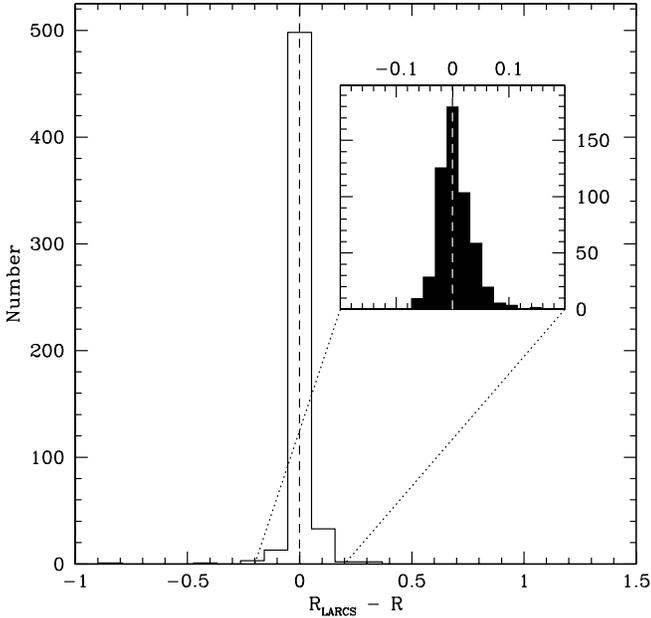,angle=0,width=3.5in}}
  \caption{\small{Histogram of the difference in R magnitude between
LARCS and this work.  The inset panel shows an enlargement of the
central region of this histogram.  The dashed vertical line is the median
offset, $\Delta R = -0.0002 \pm 0.0014$.
}}
  \label{fig:dmag}
\end{figure}

\subsection{Internal Consistency}

First, the internal consistency of the photometric solution is
tested by employing the overlap of our target observations (Figure~\ref{fig:layout}).
We use SExtractor's {\sc mag\_best}, a maximum search 
radius of $10''$ and only those sources whose {\sc flag} parameter is equal to 0,
thus eliminating saturated and heavily blended objects as well as those
that fall close to the chip edges. 
This results in 681 matched objects whose median magnitudinal difference 
from the Abell~1079 pointing to the Abell~1084 one is
$<< 0.04$ mags in all three passbands.
Further, the scatter as a function of magnitude is 
consistent with the errors given by SExtractor.
Therefore we consider the magnitudinal zeropoint error of 0.04
mags to be our dominant source of error.

As the final step, we now remove all duplicate objects identified 
in the overlap area by averaging their critical parameters (e.g.\ magnitudes)
to create our master catalogue.

\subsection{Comparison to LARCS}

The LARCS data used consists of high quality 
CCD images of a 2 degree field centred around Abell~1084 that were
obtained at Las Campanas Observatory using the 1-m Swope Telescope.  
More details of the observations, reduction and analysis of
these data are given by Pimbblet et al.\ (2001).
It is sufficient for our discussion to note that LARCS covers all of
the In-FOG-Pro observations (Figure~\ref{fig:layout}) and that
for their brighter objects ($R < 19.0$), the LARCS
internal magnitudinal errors are certainly better than $0.03$ mags.
Their pixel scale ($0.696''$ pixel$^{-1}$), however, 
is much poorer than ours ($0.2295''$ pixel$^{-1}$).
By accounting for Galactic reddening (Schlegel, Finkbeiner \& Davis 1988)
we are readily able to compare our magnitudes to the LARCS ones.

Figure~\ref{fig:dmag} displays the results of comparing all 
sources with a magnitude cut of $R_{LARCS}<19.0$ and a maximum
search radius of $3''$.  
This results in 569 matched objects.
The median magnitude offset is measured as $\Delta R = -0.0002 \pm 0.0014$ from
LARCS to this work.  Such a small value is well inside our
assumed zeropoint error of 0.04 mags.

\vspace*{0.2in}

\section{Filament Verification}

\subsection{Defining the Filament}

All clusters of galaxies display a colour-magnitude relation (CMR; Bower,
Lucey \& Ellis 1992; Visvanathan \& Sandage 1977) 
of early-type (i.e.\ elliptical and lenticular) galaxies.  
Indeed, so ubiquitous is this relation that Gladders \& Yee
(2000) are using it to discover new high-redshift clusters.
Therefore we assume that any filament that connects our two clusters
should also have early-type galaxies that lie close to the cluster's
CMR.  Moreover, galaxies lying on the CMR will be at approximately
the same redshift.

\begin{figure}
\centerline{\psfig{file=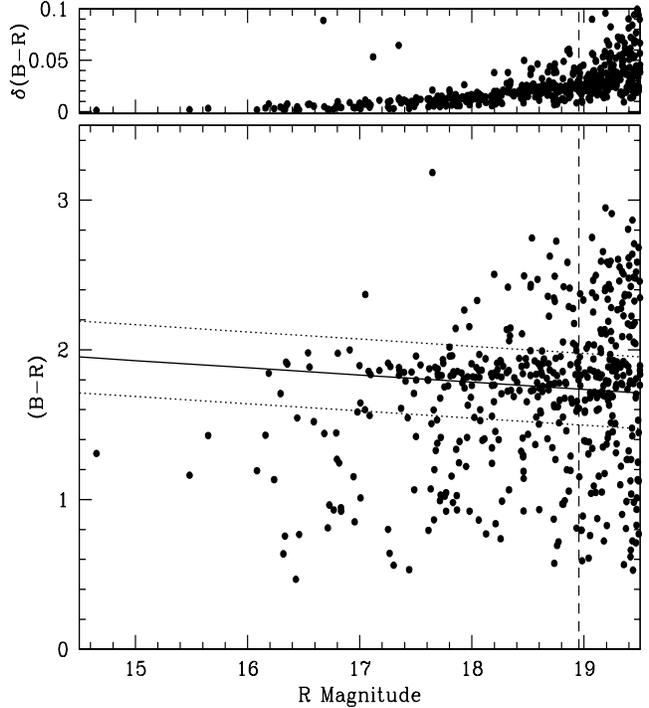,angle=0,width=3.5in,height=4.0in}}
  \caption{\small{Colour magnitude diagram for all galaxies in our
catalogue (lower panel).  Overplotted is the mean CMR for our clusters 
from Pimbblet 
et al.\ (2002), flanked by the $3\sigma$ uncertainty in 
colour (dotted lines), and $M_V=-20$ (dashed line;
derived from assuming an elliptical galaxy with $(B-R)=1.5$
evolved to zero redshift).
The upper panel displays the error in $(B-R)$.
}}
  \label{fig:cmr}
\end{figure}

Figure~\ref{fig:cmr} shows the CMR for all galaxies in our sample.
Pimbblet et al.\ (2002) have already shown that the CMR 
of Abell~1084 is very similar to that of Abell~1079.
Therefore, we make use of the mean CMR slope and intercept values published by
Pimbblet et al.\ (2002) for our two clusters; slope=$-0.048 \pm 
0.010$, intercept
$(B-R)_{M_V=-21.8}=1.815 \pm 0.079$ where $R_{M_V=-21.8}=17.155$
\footnote{Our notation $(B-R)_{M_V=-21.8}$ refers to the
$(B-R)$ colour of the CMR at an absolute magnitude of $M_V=-21.8$.
Similarly, $R_{M_V=-21.8}$ is the $R$ magnitude that
corresponds to an absolute magnitude of $M_V=-21.8$.
Further, the value of $M_V=-21.8$ has been chosen as it approximates
well the magnitude of an $L^*$ galaxy (Pimbblet et al.\ 2002).
Apparent magnitudes are computed using differential $k$-corrections
(i.e.\ a `no-evolution' model; Pimbblet et al.\ 2002; Smail 
et al.\ 1998; Kodama \& Arimoto 1997).}.
Assuming that the CMR stretches between these clusters, we are able to 
make an initial pass at defining  
filament membership (at least for the old elliptical galaxies) by 
taking galaxies within the $3\sigma$\footnote{We employ a $3\sigma$ scatter
in the $(B-R)$ colour as: (i) the CMR fit made by Pimbblet et al.\ (2002) is
only valid in the central 2 Mpc region and (ii) the width of the CMR 
is known to significantly broaden with clustocentric radius
due to age effects
(Pimbblet et al.\ 2002; Kodama, Bower \& Bell 1999).}
$(B-R)$ colour (measured with {\sc mag\_iso}) scatter of this relation
(see Figure~\ref{fig:cmr}).

%
%

\subsection{Spatial Distribution}

%
%
\begin{figure*}
\centerline{\psfig{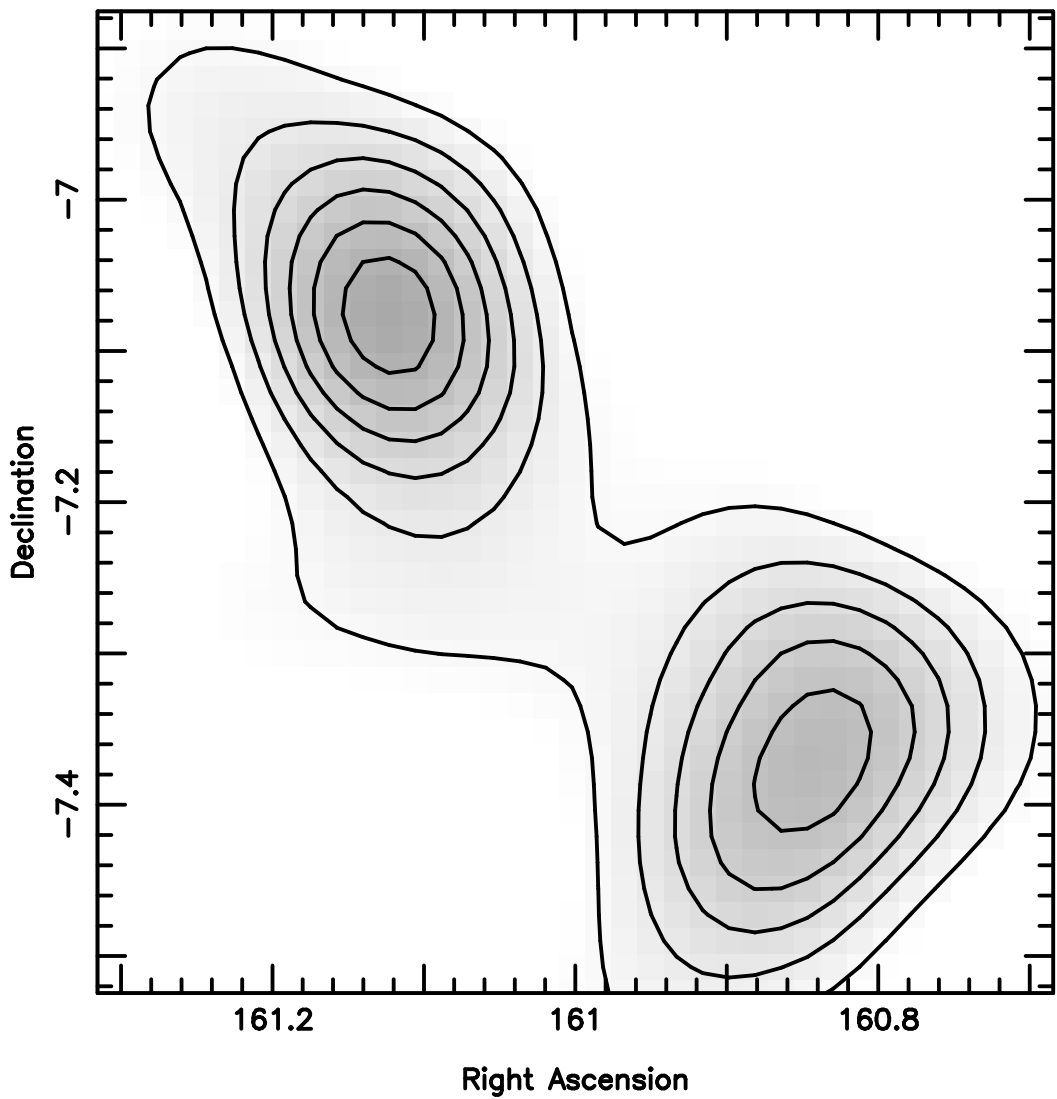}}
  \caption{\small{Smoothed spatial distribution of those galaxies 
brighter than $M_V=-20$ within the $3\sigma$ scatter of the mean 
cluster CMR.  
The positions of the galaxies are smoothed using a circular top hat
function with a smoothing length of $\approx 800$ kpc
at the mean inter-cluster redshift.  
The lowest contour is a galaxy surface density of 3 Mpc$^{-2}$
and this increases by 1 Mpc$^{-2}$ per contour inward.
The overdensities are Abell~1084 (upper left) and Abell~1079 (lower right).
Note the low density filament of early-type galaxies
that stretches between the two clusters.
}}
  \label{fig:spatial}
\end{figure*}

%
%
\begin{figure}
\centerline{\psfig{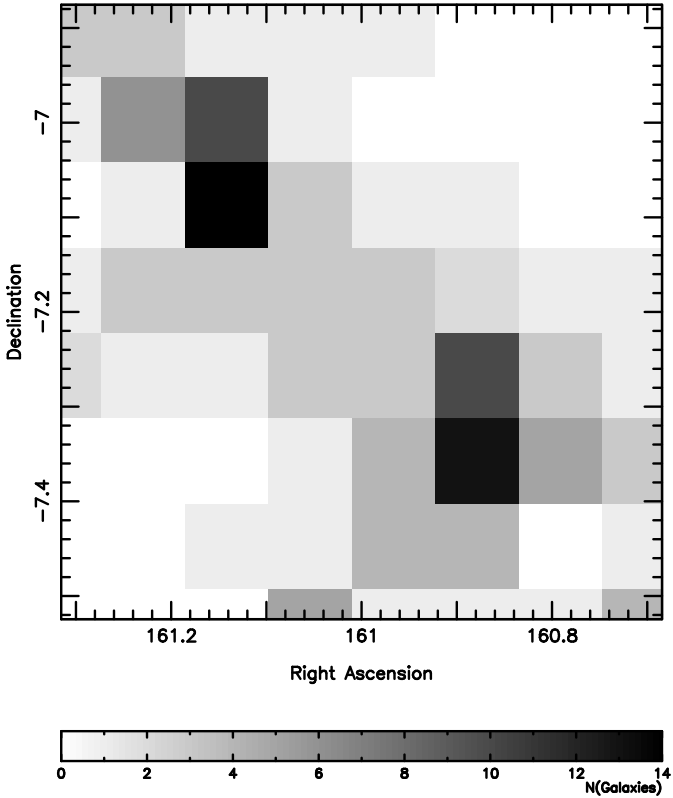}}
  \caption{\small{As for Figure~\ref{fig:spatial}, but
for a plain binned spatial distribution.  
}}
  \label{fig:spatial2}
\end{figure}

Using a fiducial limiting magnitude of $M_V=-20$
to be consistent with other studies (Butcher \& Oemler 1984; 
Pimbblet et al.\ 2002), we plot the smoothed spatial
distribution of these filament members in Figure~\ref{fig:spatial}.
For comparison purposes we provide an un-smoothed version 
of this in Figure~\ref{fig:spatial2}.
The overdensities seen are the cluster centres.
Importantly, there is a population of CMR members that
stretches between the clusters at a distance of $>3 r_{Abell}$
from either cluster (Abell, Corwin \& Olowin, 1989).  
Given the low density of these galaxies (an order
of magnitude less dense in the filament than in the cluster), 
we infer that early-type
galaxies do not dominate the filament outside the cluster centres.
Moreover, it is near this density regime ($\sim 1$ galaxy
Mpc$^{-2}$) where a 
noted turn-over in star-formation rate occurs 
(G{\' o}mez et al.\ 2003; Lewis et al.\ 2002).
We must therefore account for the bluer filament members.

\section{Colour-colour Analysis}

In order to proceed to analyze the total filament population in the absence 
of spectroscopy, we must employ some kind of statistical background
correction to remove contaminating foreground and background galaxies.
To sample the contaminating field galaxies, the two CCD chips at opposite 
extrema of the observations are used.
Using this sample will inevitably result in some small amount of 
filament galaxies being subtracted, the effect being
more pronounced at fainter magnitudes (Smail et al.\ 1998)
However, we consider this number
to be small as these CCD chips are sufficiently far away from 
our filament ($\sim 3$--$4$ Mpc) as to not contain many filament or
cluster members (Figures~\ref{fig:layout} and~\ref{fig:spatial}; also 
see Kodama \& Bower 2001 and Paolillo et al.\ 2001
for a perspective on sampling the background close to clusters).

We proceed by dividing up our filament sample into three
sub-samples: Abell~1079, Filament and Abell~1084.
These sub-samples are centred upon the clusters (Table~\ref{tab:clusters}) 
and the approximate filament centre ($\alpha=161.0$, $\delta=-7.24$; 
see Figure~\ref{fig:spatial}) out to a radius of $0.1$ degrees.
At this radius, the samples touch but do not overlap one another.

\begin{figure*}
\vspace*{-1.2in}
\centerline{\psfig{file=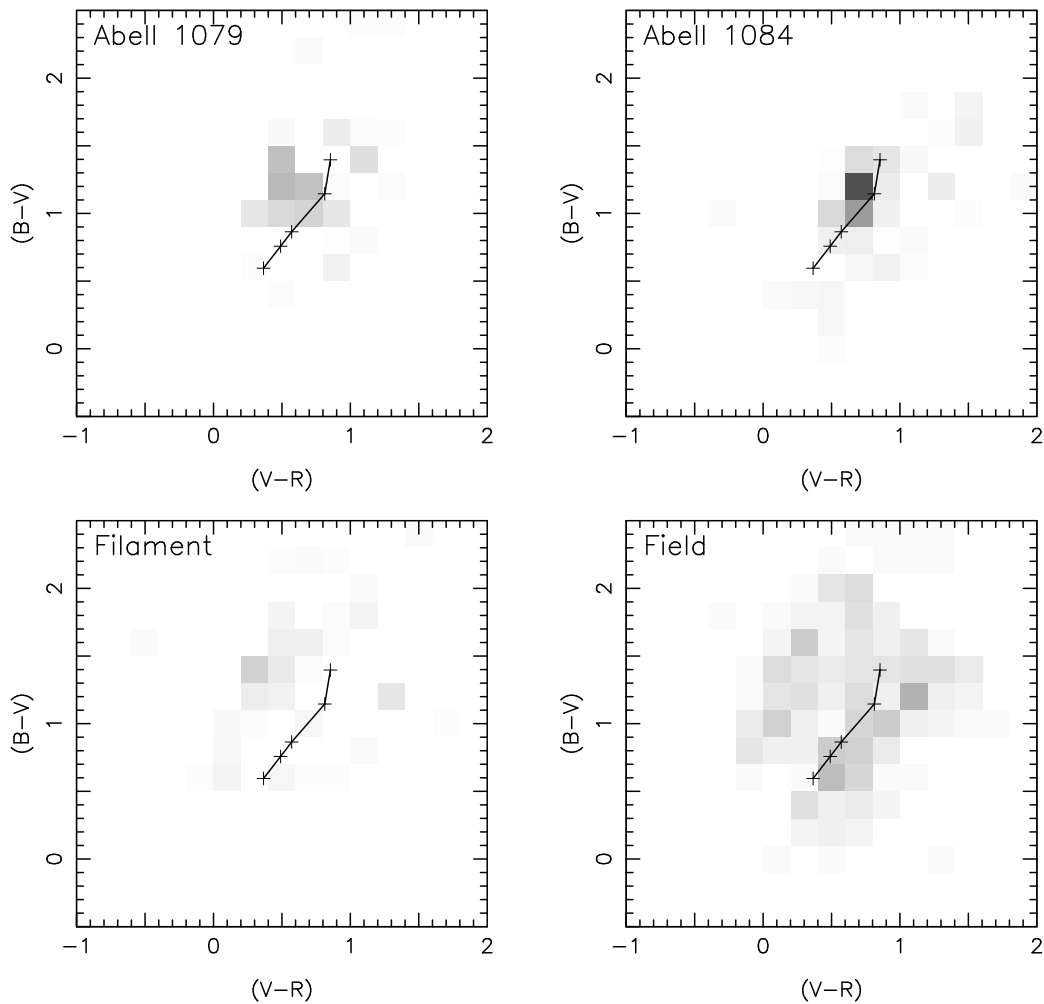,angle=0,width=7.in}}
\vspace*{-3.2in}
  \caption{\small{Colour-colour plots
of the two clusters and filament region for galaxies brighter than
$R=R_{M_V=-18}$ after correction for field contamination.
The equivalent distribution for the field sample is also shown
(lower right panel; arbitrarily scaled).
Overplotted is the locus of expected colours for non-evolving
spectral energy distributions that correspond to galaxies of 
spectral type E, Sab, Sbc, Scd and Sdm (upper right through lower left)
morphology in the local Universe as they would be observed at $z=0.1335$
(see Smail et al.\ 1998; Kodama \& Arimoto 1997; Fukugita, Shimasaku, \& 
Ichikawa 1995).
}}
  \label{fig:cc}
\end{figure*}

Colour-colour diagrams in the $(V-R)-(B-V)$ plane are constructed 
for the field and three sub-samples with $R<R_{M_V=-18}$.  
An area-scaled amount of the field sample\footnote{The area of the
field sample is $\sim 245$ arcmin$^2$.} is then subtracted 
from corresponding regions on the $(V-R)$--$(B-V)$ plane of the 
filament sub-samples to generate the final distribution.
The result of this analysis is presented in Figure~\ref{fig:cc}.
We overlay on these plots the expected colours of various local 
morphological types, assuming a no-evolution model, as 
they would be observed at the mean redshift of the clusters.

%
%

\subsection{Discussion}

Both of the clusters show a clear overdensity of early- to mid-type 
galaxies with a number of bluer galaxies.  Given the cluster
blue fractions ($f_B$) calculated by Pimbblet et al.\ (2002), this
is unsurprising.
Conversely, the filament's galaxian population is an ecletic mixture of
bluer galaxies that span a range of star formation rates 
(Figure~\ref{fig:cc}).

In order to better constrain the existence of the filament,
we now estimate its overdensity from an analysis of Figure~\ref{fig:cc}.
Since we know the expected field population ($N_{field}$), 
we can compute the excess number of galaxies
above this level for the filament sub-sample on the colour-colour
plane (akin to Pimbblet et al.\ 2002).  
Following Paolillo et al.\ 2001, the galaxy number excess
is given by: 

\begin{equation}
N_{filament} = N_{filament + field} - N_{field}
\end{equation}

If our field sample has been chosen too close to the clusters,
however, it will obviously contain some (small but
non-negligible) amount of contamination:

\begin{equation}
N_{field}^{'} = N_{field} + \gamma N_{filament}
\end{equation}

where $\gamma$ is simply the ratio between the galaxy densities of the
filament and field populations (Paolillo et al.\ 2001).
Substituting $N_{field}^{'}$ instead of $N_{field}$ from
Eq.\ (2) into Eq.\ (1) generates

\begin{equation}
N_{filament}^{'} = N_{filament} (1 - \gamma)
\end{equation}

\begin{figure*}
\vspace*{-0.6in}  
\centerline{\psfig{file=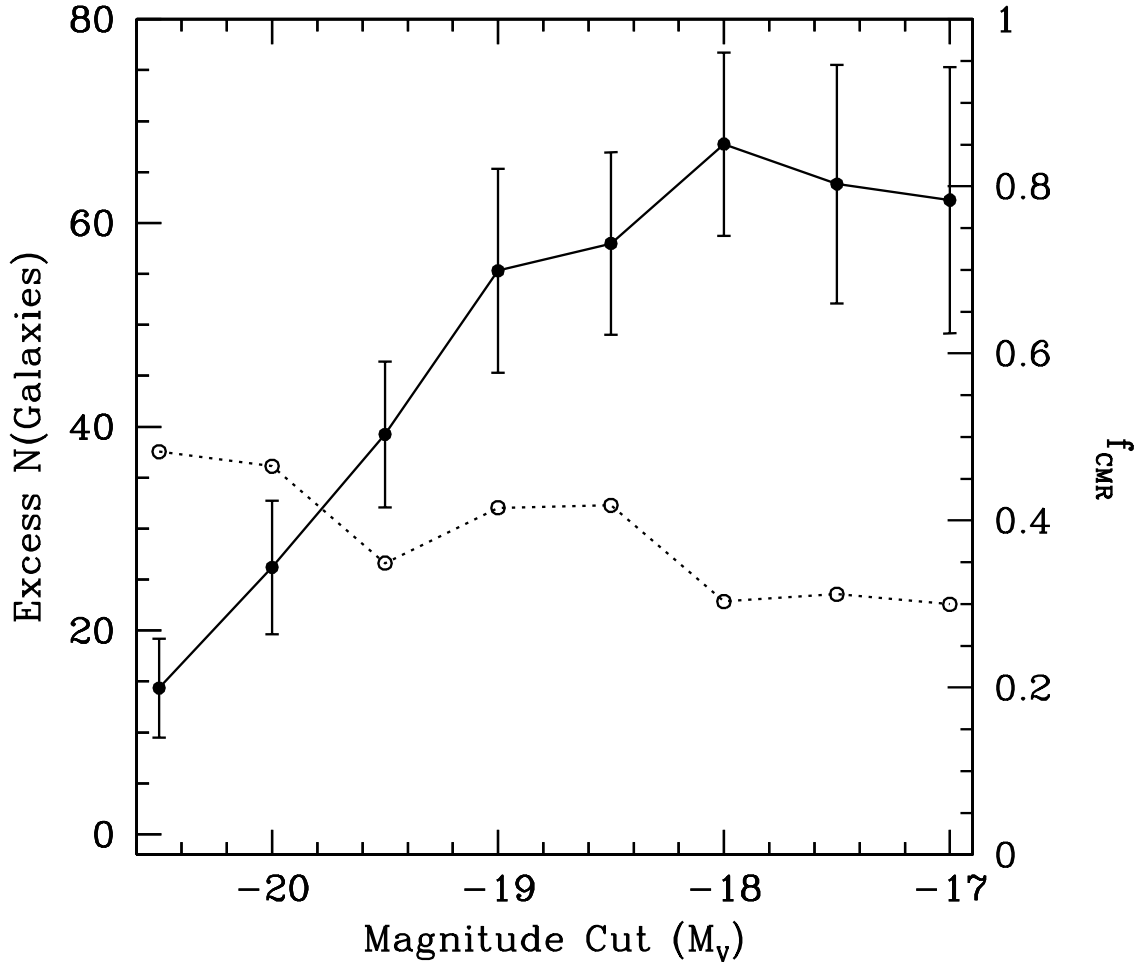,angle=0,width=7.in}}
\vspace*{-0.4in}  
\caption{\small{Number of galaxies in excess of the field
population for the filament sub-sample as a function of
limiting magnitude cut (filled circles; solid line).  
The $\pm 1 \sigma$ errorbars come from a consideration
of Poissonian error and galaxy number density variance.
Overplotted is the fraction of these galaxies that lie on the
CMR (i.e.\ possess colours that are consistent with the $3\sigma$
uncertainty of the mean CMR; Figure~\ref{fig:cmr}), $f_{CMR}$, (open circles;
dotted line) and are thus expected to be at the filament redshift.
}}
  \label{fig:xs}
\end{figure*}

By using a field sample too close to our filament sub-sample
this implies that our
final number counts will be diminished somewhat.
It follows that 
the ultimate error budget is therefore composed of a
number of sources: Poissonian errors for the filament count,
Poissonian errors for the field and galaxy number
density variance.
Since the area of the field sample is $\gg$ than the area of the 
filament sample, the field Poissonian error is small and the
other two sources of error dominate the final error.
The galaxy number density variance is measured directly from
the catalogues whilst Poissonian errors are computed from the
field correction technique.  
The number of galaxies found in excess of the field 
population is plotted in Figure~\ref{fig:xs}.
Thus at a magnitude of $M_V = -18$, the filament is
detected at a $\sim 7.5 \sigma$ level (falling to $5.5 \sigma$
at $M_V=-19$ and $4 \sigma$ at $M_V=-20$; Figure~\ref{fig:xs}).
We are also able to compute the fraction of these galaxies that
have colours consistent with the CMR, $f_{CMR}$.
Significantly, $f_{CMR}$ gives us an approximate 
yet effective lower bound
to the number of galaxies expected to be at the filament
redshift (since CMR members are early-type galaxies at 
similar $z$).
This fraction ranges from $f_{CMR} = 0.45$ at bright magnitudes to 
$f_{CMR} = 0.30$ at fainter ones (Figure~\ref{fig:xs}).

\begin{figure*}
\vspace*{-1.2in}
\centerline{\psfig{file=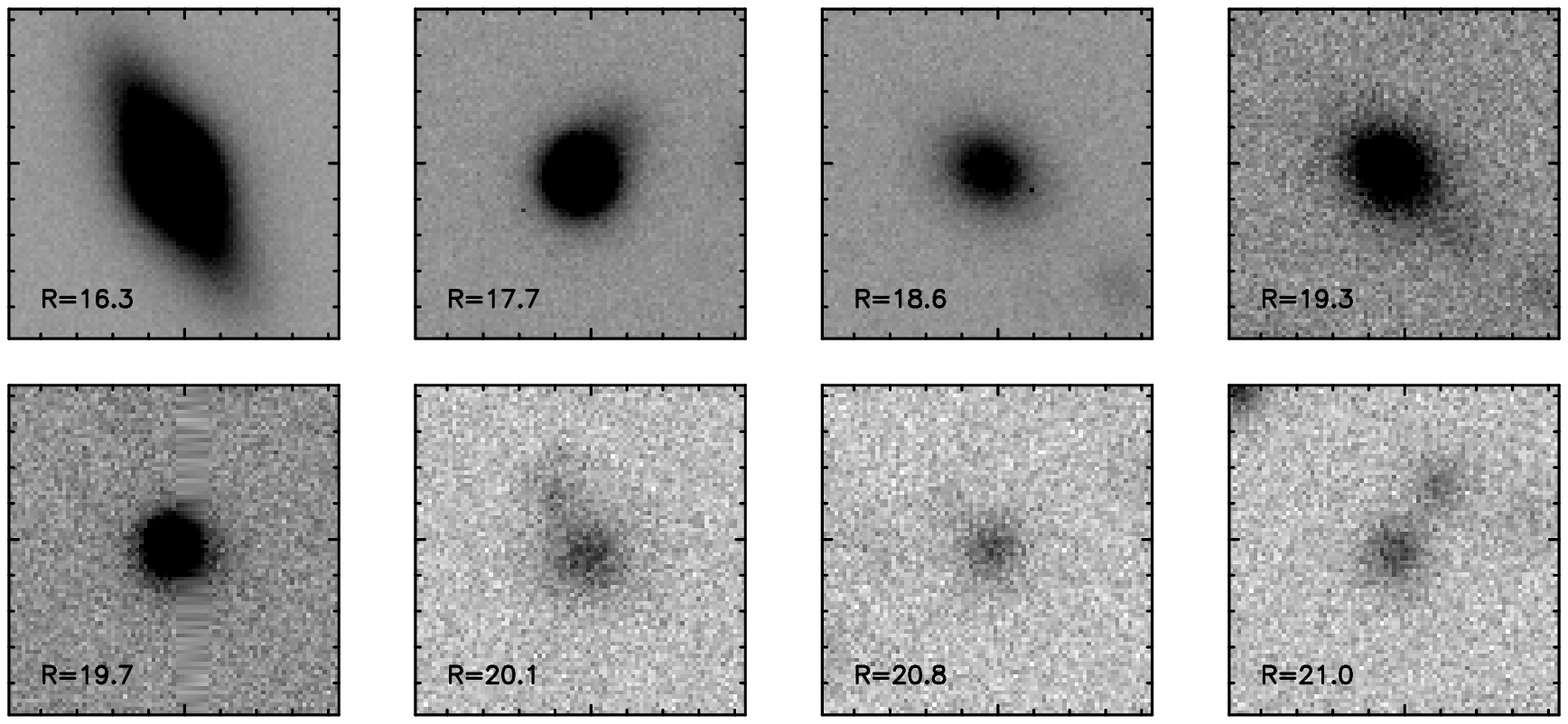,angle=0,width=7.in}}
\vspace*{-5.8in}
  \caption{\small{Examples of prospective filament members based
on their $(V-R)$ and $(B-V)$ colours from bright to faint
magnitude (noted at the lower left).  
Each image is taken from one of the $R$--band CCD images
of our observations.  They are $20''$ on the side, each tick
mark thus equivalent to $2''$.  At the mean redshift of the clusters,
this makes each thumb-nail image approximately $60$ kpc on the side.
}}
  \label{fig:eg}
\end{figure*}

We now attempt to visually characterize the filament population
by selecting filament galaxies whose colours make
them part of the $f_{CMR}$ fraction
and are hence highly likely to be at the filament redshift\footnote{We 
emphasize the caveat that there will be some small
amount of contamination in the CMR population.  Given their
colours, we are confident that the majority of these potential 
filament members are at the appropriate redshift.}.
Figure~\ref{fig:eg} displays a selection of these {\it potential}
filament galaxies.
Whilst there are many of early-type galaxies (e.g.\ $R=19.7$, 
Figure~\ref{fig:eg}), some of them
are morphologically mid-type galaxies (e.g.\ $R=16.3$, 
Figure~\ref{fig:eg} is visually classified as an Sab type).
A couple possess features indicative of recent interactions
(e.g.\ the $R=17.7$ galaxy in Figure~\ref{fig:eg} has a distinct spike to 
its upper right whilst the $R=20.1$ galaxy appears to have either just
undergone an interaction with its neighbour to its upper left, or is 
just about to).

It is these galaxies that will eventually be accreted to the clusters.
Although on the filament and above the critical turn-over for
star-formation-density (3--4 Mpc$^{-2}$ c.f.\ 1--2 Mpc$^{-2}$; 
G{\' o}mez et al.\ 2003; Lewis et al.\ 2002),
some of these galaxies appear to still be actively star-forming
(Figures~\ref{fig:cc} and~\ref{fig:eg}). 
Therefore, whatever the environmental 
mechanism for turning off star-formation
in the central regions of clusters (e.g.\ Pimbblet 2003),
it has affected some filament galaxies and not others.
Given that the present study only samples one filament,
further studies into this density regime are urgently required to
differentiate between the possible mechanisms.  
If the strangulation scenario forwarded by Balogh et al.\ (2000)
is confirmed, then it {\it must} be able to operate in these
low density environs (see also Balogh et al.\ 2002;
Bekki, Couch and Shioya 2001; Fasano et al.\ 2000).
Thus it is in studies such as the present one that
we will be able to observe the onset of strangulation.

%
%

%
%

\section{Summary}

This work has presented the multicolour (BVR) CCD pilot 
observations for a larger study
of inter-cluster filaments of galaxies.
Inparticular, we have shown that:

\begin{itemize}

\item A low density filament of galaxies stretches between 
Abell~1079 and Abell~1084 and we have detected it at
a $7.5 \sigma$ level ($M_V=-18$ cut). 
\item The early-type component of this filament is 
an order of magnitude less dense than similar ones in the cluster
centres, perhaps 3--4 galaxies per Mpc$^2$.
\item The bluer galaxies are mostly morphologically mid- to early-types
with a number displaying signs of recent interactions.
This point will be examined in more detail in a forthcoming
article.
\item Given the colours and states of these filament galaxies,
whatever the mechanism responsible for suppressing star-formation rate
in cluster centres, it is only just beginning to have an effect in
the filamentary region.

\end{itemize}

This is the first publication in a series
based on the In-FOG-Pro observations.
In our future publications we will be able to combine observations from
numerous filaments resulting in significant improvement of our signal to noise 
and hence be able to better characterize the galaxy populations
in such low density environments.
Using our planned spectroscopic followup observations we will also be able
to examine star formation rates using spectral lines (e.g.\ [OII]; H$\delta$)

\section*{Acknowledgements}
The imaging observations presented in this work were taken in WFI service mode
operations at the AAT in February 2003 by C.\ Tinney whom we wish to
thank.
We also thank useful conversations with A.\ Karick, G.\ Dalton,
I.\ Smail, A.\ Edge and W.\ Couch.
M.\ Hawkrigg is also thanked for her careful reading of this
work which has significantly improved its quality.
KAP acknowledges support from an EPSA University of Queensland Research
Fellowship.

\end{document}